\documentclass[%
 amsmath,amssymb,
 reprint,%
]{revtex4-1}

\usepackage{graphicx}% Include figure files
\usepackage{dcolumn}% Align table columns on decimal point
\usepackage{color}
\usepackage{bm}% bold math
\begin{document}

%\preprint{AIP/123-QED}

\title{Resistive Switching in Memristive Electrochemical Metallization Devices}% Force line breaks with \\

\author{Sven Dirkmann}
\email{sven.dirkmann@rub.de}
 \affiliation{Institute of Theoretical Electrical Engineering, Ruhr University Bochum, D-44780 Bochum, Germany. }%Lines break automatically or can be forced with \\
\author{Thomas Mussenbrock}%
\affiliation{Electrodynamics and Physical Electronics Group 
Brandenburg University of
Technology, D-03046 Cottbus, Germany.%\\This line break forced with \textbackslash\textbackslash
}%

\date{\today}% It is always \today, today,

\begin{abstract}
We report on resistive switching of memristive electrochemical metallization devices using 3D kinetic Monte Carlo simulations describing the transport of ions through a solid state electrolyte of an Ag/TiO$_{\text{x}}$/Pt thin layer system. The ion transport model is consistently coupled with solvers for the electric field and thermal diffusion. We show that the model is able to describe not only the formation of conducting filaments but also its dissolution. Furthermore, we calculate realistic current-voltage characteristics and resistive switching kinetics. Finally, we discuss in detail the influence of both the electric field and the local heat on the switching processes of the device.
\end{abstract}
% We added this reference to the manuscript.
%Use showkeys class option if keyword
\maketitle

\section{Introduction}

Memristive devices are devices which change their resistance by applying a voltage and maintain this value when removing it. This is actually the basis of resistive switching.  Due to their unique key features, including an excellent miniaturization potential ($<10$ nm), high operation speed, low energy consumption ($< \text{pJ}$) and high endurance ($>10^{12}$ switching cycles), memristive devices have attracted a lot of attention as potential future non volatile memories and as artificial synapses within neural networks \cite{Pan:2014, Hansen:2017}. Electrochemical metallization (ECM) cells are in principle such memristive devices. They consist of a low conductive layer, which is an ionic/electronic mixed conductor, sandwiched between a chemically active and an inert metal electrode \cite{Waser:2007}. When a positive voltage is applied to the active electrode, metal atoms can oxidize and drift and/or diffuse through the low conductive layer towards the inert electrode. Here, stable nuclei can be formed and a conductive metal filament starts growing, bridging the low conductive layer and setting the device into a low resistive state (LRS). When the polarity of the applied voltage changes, the filament dissolves and the device is reset back into the high resistive state (HRS) again. 

Although the basic concept is quite simple, the underlying physical mechanism of resistive switching is highly complex. In order to gain a profound understanding, it is crucial to model the physical processes on the atomic scale. Up to now, a number of different kinds of models have been developed in order to investigate resistive switching in ECM cells. Besides compact models \cite{Menzel:2012, Yu:2011, Russo:2009}, continuum models \cite{Lin:2012} and molecular dynamics models \cite{Onofrio:2015}, also kinetic Monte Carlo (KMC) models have been reported \cite{Jameson:2011, Jameson:2012, Pan:2011, Qin:2015, Menzel:2015, Dirkmann:2015}. However, although filament formation is an inherent 3D phenomenon, a 3D model, which is able to track the dynamics on experimental length and time scales, is missing. 

The goal of this work is to provide a kinetic 3D model which allows to investigate the set and the reset of a Ag/TiO$_{\text{x}}$/Pt ECM cell. Since the KMC method offers the possibility to calculate the inner atomic state of a resistive switching device on experimental length and time scales, here all atomic and ionic processes have been calculated using this approach. Furthermore, in contrast to KMC models of ECM-cells proposed up to now, our model is coupled to the heat equation. The temperature distribution, which has been calculated from the heat equation is coupled back to the KMC model in order to allow for its influence on ionic processes. The electric field is calculated solving the continuity equation whereas the current through the device is calculated using a generalized  Ohm's law. It is shown, that the calculated IV-characteristic of the device is in good agreement with experimental findings. Ultimately, the switching kinetics of memristive ECM-cells is discussed. 
\begin{figure}[t!]
\centering
\includegraphics[width=3.3in]{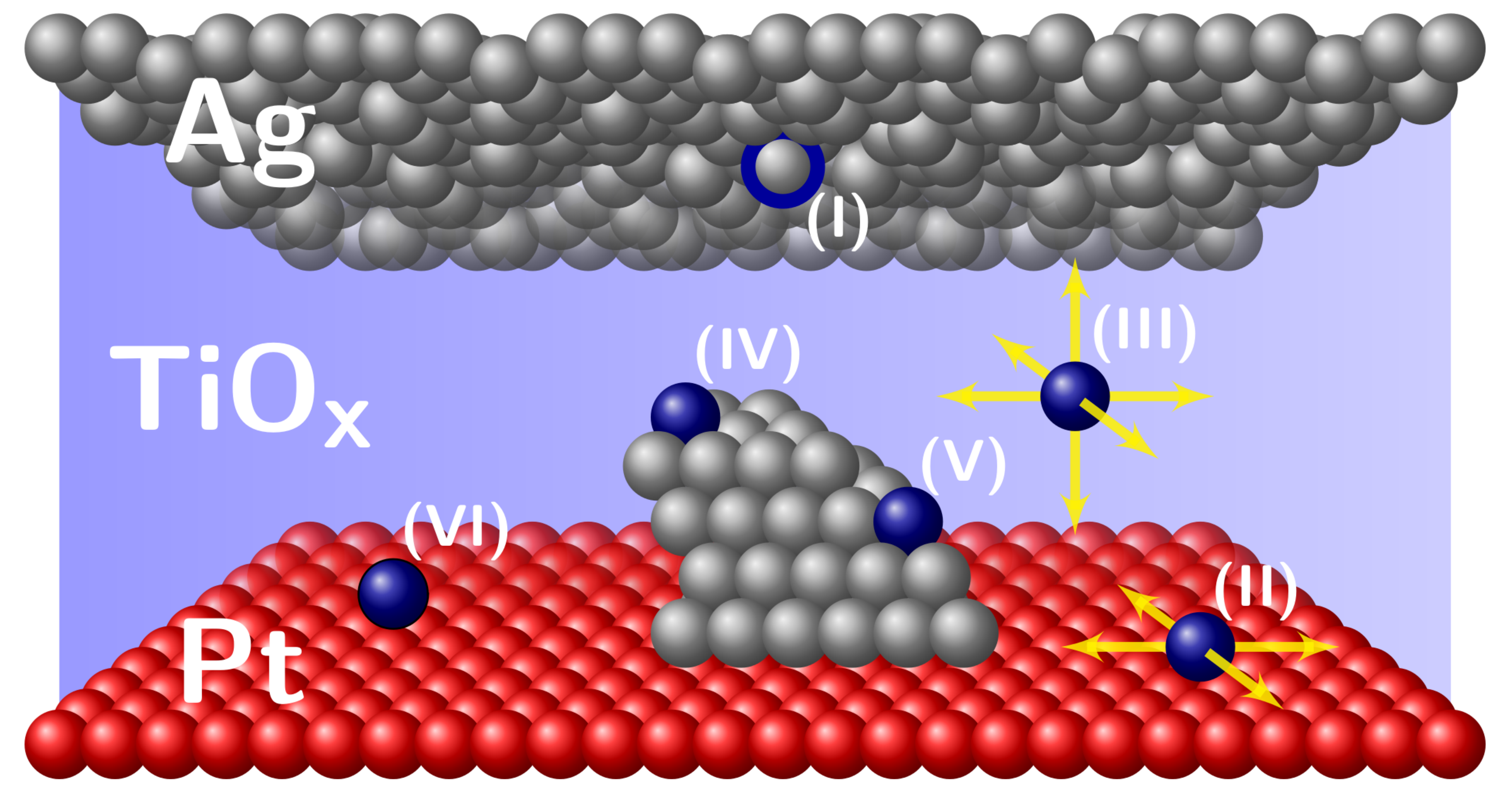}
\caption{Illustration of the processes included within the simulation: (I) oxidation, (II) surface diffusion, (III) diffusion within TiO$_{\text{x}}$, (IV) reduction at Ag, (V) reduction at kink side and (VI) nucleation.}
\label{Processes}
\end{figure} 

\section{Simulation approach}

\begin{table}[!t]

\caption{Simulation parameters}
\label{table_example}
\centering
\begin{tabular}{p{2.1in}c}

\hline 
\rule{0pt}{3.ex} Physical quantity & Value \vspace{3pt} \\

\hline 

Conductivity $\sigma$ of Ag & $6.3\cdot 10^{7}$ S/m \\
Conductivity $\sigma$ of TiO$_{\text{x}}$ & $1 \cdot 10^{2}$ S/m\\
Ionic charge number $z$ & 1\\
E$_{ox}$ oxidation of Ag & 0.65 eV\\
E$_{red}$ reduction (surface/kink) & 0.62/0.58 eV \\
E$_{nuc}$ nucleation & 0.81 eV \\
E$_{a}$ ion diffusion & 0.61 eV\\
E$_{a}$ surface diffusion & 0.59 eV\\
Mass density $\rho$ of Ag\cite{Enghag2004}/TiO$_{\text{x}}$\cite{Kharisov2016} & 10490/4230 kg/m$^3$ \\ 
Heat capacity $c_p$ of Ag\cite{Abu-Eishah:2004}/TiO$_{\text{x}}$\cite{Saeedian:2013} & 235/700 J/(kg K) \\
Th. conductivity $\lambda$ of Ag\cite{Ho:1972}/TiO$_{\text{x}}$\cite{Lu:2013}  & 429/7 W/(m K) \vspace{3pt} \\
\hline
\end{tabular}
\end{table}

The simulation domain consists of a 40 nm $\times$ 40 nm square base, a 10 nm thick TiO$_{\text{x}}$ solid electrolyte and a 3 nm thick Ag electrode (see figure \ref{AtomicState}). The Ag electrode is constituted of 38400 individual atoms (gray), whereas the Pt electrode is given just by a stopping layer at the bottom of the simulation domain. The atoms of the chemically active Ag electrode can undergo the following implemented physical processes, presented in figure \ref{Processes}: (I) Oxidation, (II) surface diffusion, (III) diffusion within the TiO$_{\text{x}}$, (IV) reduction at Ag surface and (V) kink side and (VI) nucleation. A cubic lattice with the lattice constant of 0.5 nm, has been applied. The corresponding lattice constant is equal to the hopping distance of the ions.
 
Within the KMC approach, occurring physical processes are described by rate equations. The motion of Ag ions is driven by the potential difference between to neighbouring lattice sites and the corresponding reaction rate is given by the Arrhenius law:
\begin{equation}
k_{ij} = \nu_0 \exp\left[-\frac{E_{a}+0.5ze\left(E_j-E_i\right)}{k_{\text{B}} T(\vec{r},t)}\right]
\label{1}
\end{equation}
with $\nu_0 = 10^{12}$ Hz the phonon frequency, $E_{a}$ the energy barrier of the corresponding process, $k_{\text{B}}$ the Boltzmann constant, $e$ the elementary charge, $z$ the ionic charge number and $T(\vec{r},t)$ the temperature at the atomic position. The term $0.5\left(E_j-E_i\right)$ is a linear interpolation of the potential at the lattice positions $i$ and $j$. Important simulation parameters are shown in table \ref{table_example}. The reaction rates for oxidation, reduction and nucleation are deduced from the Butler-Volmer equation and are given as \cite{Menzel:2015, Qin:2015, Ielmini:2016}
\begin{equation}
k_{\text{nuc}} = \nu_0 \exp\left[-\frac{E_{\text{nuc}}-\alpha ze \Delta \phi}{k_{\text{B}} T(\vec{r},t)}\right],
\label{2}
\end{equation}
\begin{equation}
k_{\text{red}} = \nu_0 \exp\left[-\frac{E_{\text{red}}-\alpha ze \Delta \phi}{k_{\text{B}} T(\vec{r},t)}\right]
\label{2}
\end{equation}
and
\begin{equation}
k_{\text{ox}} = \nu_0 \exp\left[-\frac{E_{\text{ox}}+(1-\alpha) ze \Delta \phi}{k_{\text{B}} T(\vec{r},t)}\right]
\label{3}
\end{equation}
Here, $E_{\text{red}}$, $E_{\text{nuc}}$ and $E_{\text{ox}}$ are the activation energies for reduction, nucleation and oxidation, respectively (see table \ref{table_example}), $\alpha = 0.5$ is the charge transfer coefficient and $\Delta \phi$ is the overpotential at the electrode/electrolyte interfaces. The overpotential at the electrode(active/passive)/TiO$_{\text{x}}$ interface is given as $\Delta\phi = \phi_{\text{TiO}_{\text{x}}} - \phi_{\text{electrode}}$ and the overpotential at the filement/TiO$_{\text{x}}$ interface is given as $\Delta\phi = \phi_{\text{TiO}_{\text{x}}} - \phi_{\text{filament}}$. Within the simulations the nucleation process has been modeled as a one particle process in order to save computational time and the slowing effect due to the need of a critical number of atoms has been realized by an enhanced nucleation activation energy \cite{Qin:2015}. The time t$_{\text{set}}$, required to set a filament completely, typically decreases exponentially with the applied voltage. This is implemented by the rate equations \eqref{1}-\eqref{3}. However, for low applied voltages, t$_{\text{set}}$ diverges toward infinity\cite{Russo:2009}. Therefore, any oxidation step of Ag atoms has been complemented by the applied voltage dependent condition $r < \log[\cosh(n \cdot V_{\text{appl}})]$ using a uniquely distributed random number $r$ ($0 \leq r \leq 1$) and the fit parameter $n$. This function and the different activation energies can be used to fit the simulation to different material systems and experimental findings.

The KMC procedure evolves then as follows: For a given internal state of the cell a subset of atoms and ions is chosen randomly and they undergo randomly chosen processes depending on the calculated reaction rates for each process. Finally, the new spatial distribution of atoms is realized. The temporal evolution is realized by enhancing the simulation time $t \rightarrow t+\Delta t$. In KMC simulations the time step $\Delta t$ is not prescribed but calculated from the maximum chosen hopping rate of the iteration, $k_{\text{max}}$, by $\Delta t=-\log(r)/k_{\text{max}}$. A detailed description of the KMC procedure can be found in \cite{Dirkmann:2015, Dirkmann:2016}.

The electric potential $\Phi(\vec{r})$ within the simulation domain has been calculated solving the continuity equation  $\nabla \cdot \vec{j} = 0$, neglecting the displacement current. The current density $\vec{j}$ has been calculated by a generalized form of Ohm's law $\vec{j} = \sigma(\vec{r})\vec{E}(\vec{r})$ depending on the local conductivity $\sigma(\vec{r})$ of the respective material and on the local electric field $\vec{E}(\vec{r})$. It is important to note that, due to the assumed comparatively high conductivity of the TiO$_{\text{x}}$, the ohmic current is many times larger than the tunneling current, even for extremely small gaps. Therefore, within this model, the tunneling current has been neglected and a purely classical model has been applied. Since the dynamics of the system can be assumed to be quasi static, the electric potential can be expressed as $\vec{E} = -\nabla \Phi$. Thus, the equation for the electric potential is given by 
\begin{equation}
\nabla \cdot \left(\sigma(\vec{r})\nabla \Phi(\vec{r})\right) = 0.
\end{equation}
In order to solve this differential equation, Dirichlet boundary conditions have been applied at the top and bottom of the simulation domain. At all other boundaries periodic boundary conditions have been applied. This elliptic boundary value problem has been solved numerically on the structured Cartesian grid using the successive over relaxation method. The current $I$ through the device has been calculated by solving the surface integral 
\begin{equation}
I = \int_A \sigma(\vec{r})\vec{E}(\vec{r})\cdot \vec{n} da
\end{equation}
which can be calculated at an arbitrary vertical position since $\nabla \cdot \vec{j} = 0$ is valid.

All rate coefficients of the KMC procedure depend exponentially on the local temperature. Therefore, the temperature is a crucial parameter for resistive switching. When the ECM cell is in the LRS, a metallic filament with a diameter in the order of nm connects the top and bottom electrode. In that case, a huge current is flowing through this metallic bridge possibly leading to a significant temperature increase due to Joule heating. Thus, the assumption of a constant temperature within filamentary memristive devices might be inadequate. In this model the local temperature is not set constant, but calculated by solving the heat transfer equation
\begin{equation}
c_p(\vec{r})\rho(\vec{r}) \frac{\partial T(\vec{r},t)}{\partial t} - \lambda(\vec{r}) \nabla^2 T(\vec{r},t) = \frac{j^2(\vec{r},t)}{\sigma(\vec{r})}.
\end{equation}
Hence, the local temperature $T(\vec{r},t)$ is calculated from the current density $j(\vec{r},t)$ and the local conductivity $\sigma(\vec{r})$. $c_p(\vec{r})$, $\rho(\vec{r})$ and $\lambda(\vec{r})$ are the material-dependent heat capacity, mass density and thermal conductivity. Initially, the device temperature is set to 300 K. Furthermore, the top and bottom of the device are set to a fix temperature of 300 K during the whole simulation. At all other boundaries periodic boundary conditions have been applied. This partial differential equation has been discretized in time and on the grid, used for the KMC procedure and has been solved numerically using the single-step-method. Next, this local temperature has been inserted into eq. \eqref{1}-\eqref{3} and is therewith used to calculate the atomic evolution of the system.

\section{Results and discussion}

\begin{figure}[t!]
\centering
\includegraphics[width=7.5cm]{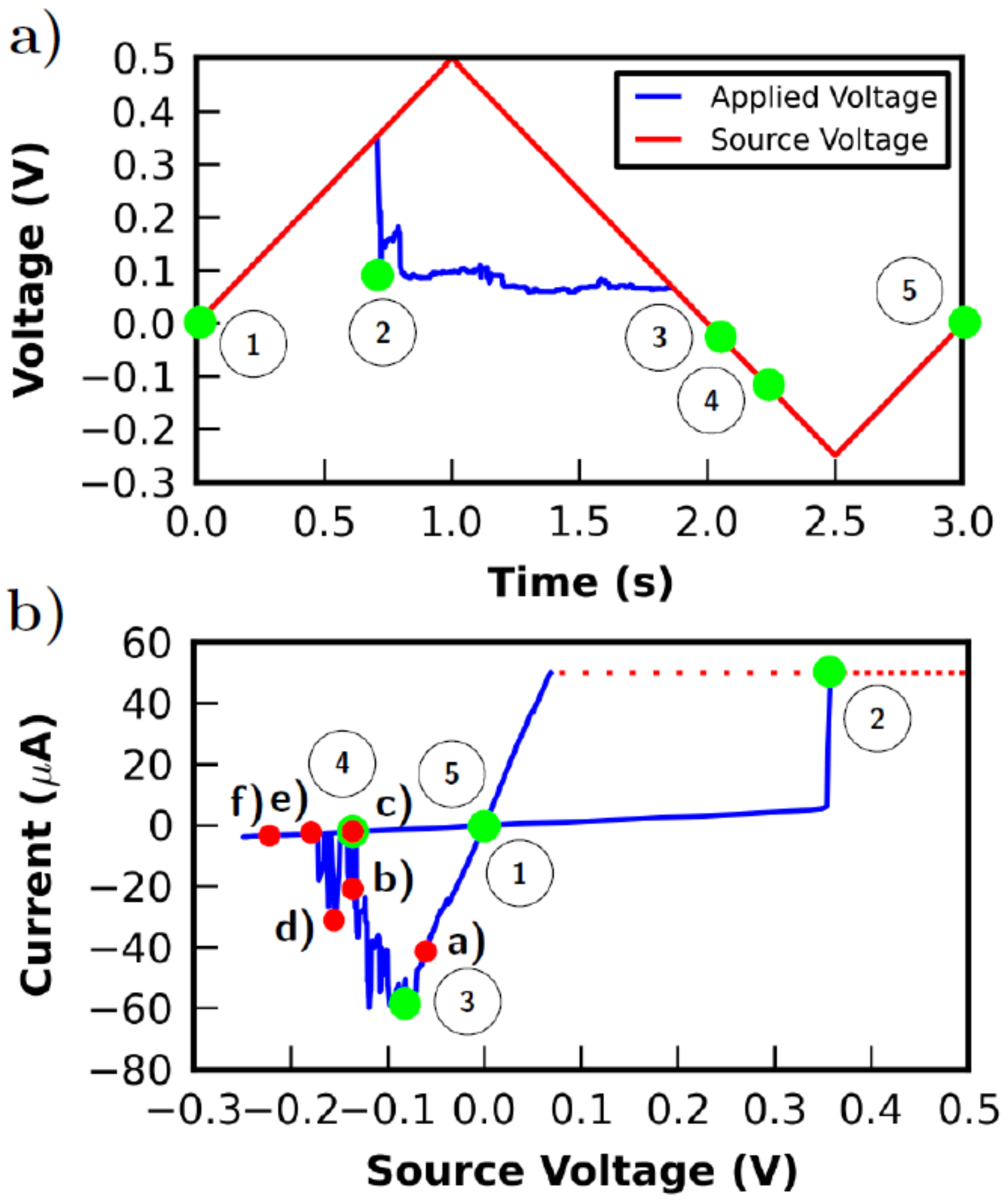}
\caption{Input and Output behaviour of the ECM cell. a): Ramp voltage of the voltage source (red) and applied voltage to the simulation domain (blue) over simulated time; b): IV-characteristic over source voltage.}
\label{IV}
\end{figure}

\begin{figure}[t!]
\centering
\includegraphics[width=3.37in]{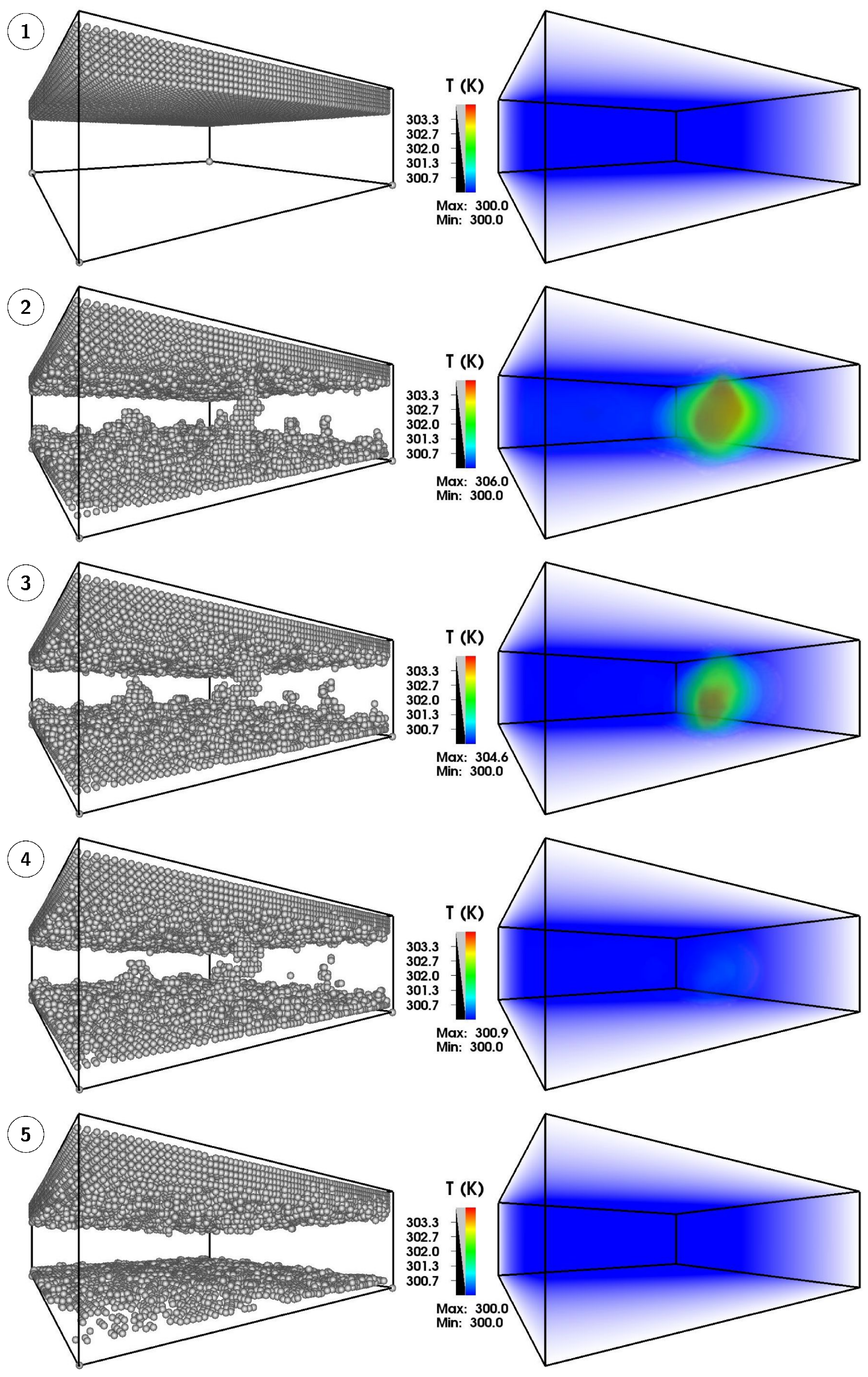}
\caption{Sequence of states at five different instants of time. Left: Status of the filament growth; right: Corresponding temperature distribution (K).}
\label{AtomicState}
\end{figure}

The presented model has been used to calculate the IV-characteristic of the ECM cell as well as the switching kinetic and the inner atomic state. An ideal voltage source has been applied to the top electrode and the bottom electrode is set to a constant potential of 0 V. The source voltage, the applied voltage to the device and the calculated IV-characteristic are shown in figure \ref{IV}. The calculated IV-characteristic is comparable to typical IV-characteristics of ECM-cells (cf.\cite{Yang:2011}). Five important time steps are marked by the numbers (1)-(5) and six different time steps of the reset process are marked by the labels a)-f). The ideal voltage source provides a voltage ramp of 0.5 V/s. In order to prevent a disruptive breakdown of the device, the maximum current is set to the compliance current of 50 $\mu$A. When the current overcomes the compliance current, the electronic excitation is switched to an ideal current source. Under this condition, the applied voltage to the device is calculated from the compliance current as well as from the actual resistance and differs from the source voltage. When the source voltage reaches the value of 0.5 V (which has been chosen to ensure resistive switching), the voltage ramp is switched to -0.5 V/s. Once the current becomes smaller than the compliance current, the current source is re-substituted by the voltage source. When the source voltage reaches the value of -0.25 V the voltage ramp is switched again to the value of 0.5 V/s until 0 V is reached. The calculated temperature distribution and the associated inner atomic state for the selected time steps, presented in figure \ref{IV}, are shown on the right hand side and left hand side of figure \ref{AtomicState}, represented by the numbers (1) - (5).
 
\begin{figure}[h!]
\centering
\includegraphics[width=3.3in]{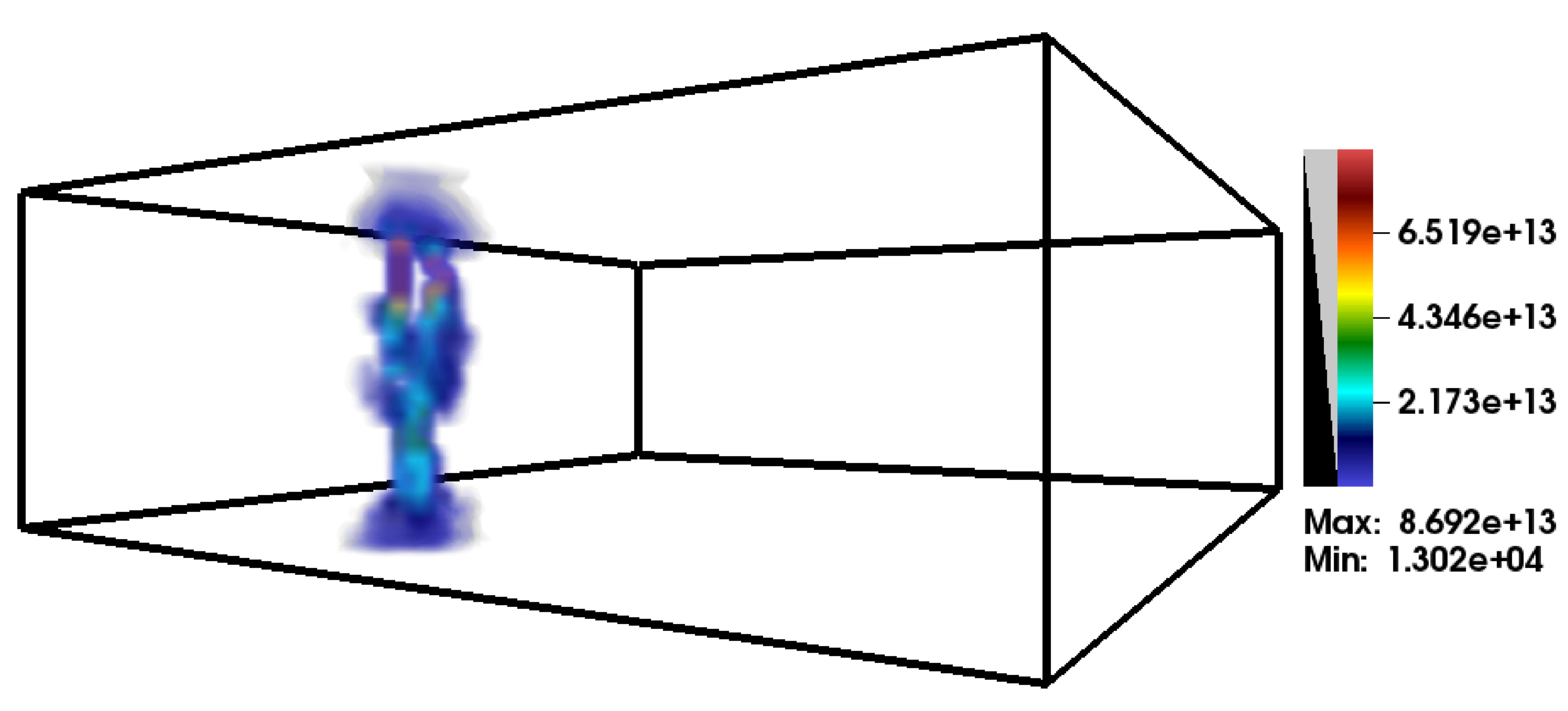}
\caption{Calculated current density (A/m$^2$) at time step (2) right after filament formation.}
\label{CurrentDensity}
\end{figure}

\begin{figure}[h!]
\centering
\includegraphics[width=3.3in]{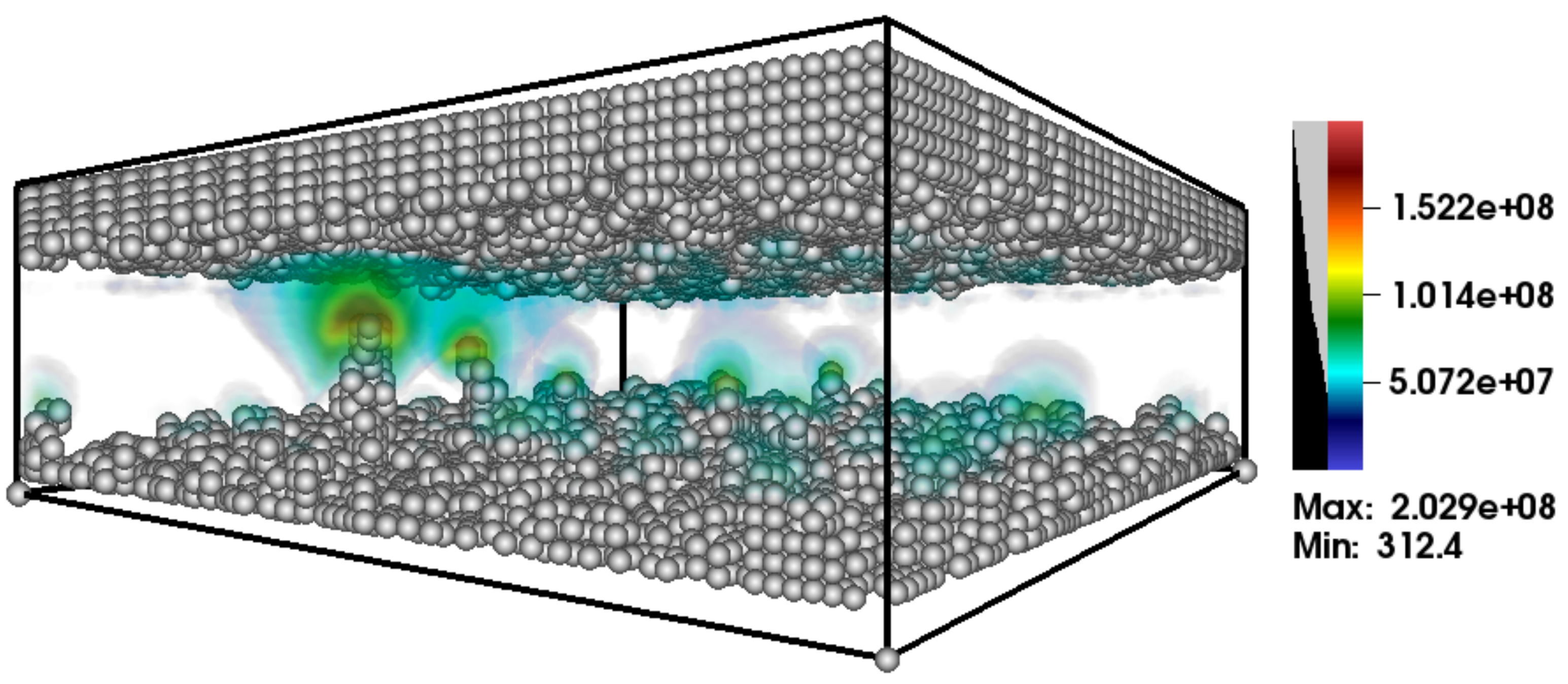}
\caption{Calculated electric field (V/m) at the instant of time t = 0.66 s right before filament formation.}
\label{Field}
\end{figure}

Initially, when the applied voltage is zero, the current is zero too and all Ag atoms are located within the electrode (1). The temperature within the device is at room temperature. When a positive voltage is applied to the Ag electrode, the electrode starts to oxidize due to the internal electric field. Ag ions move within the electric field towards the opposite electrode and the filament starts to grow. Once the Ag filament establishes a galvanic connection to the Ag electrode, a significant current starts to flow through the device (2). It is important to note, that also the remaining of an ultra thin gap between electrode and filament is possible. This can also be depicted by this model. The remaining of a gap leads to higher cell resistances than a galvanic contact and therewith to smaller temperatures due to Joule heating within the device for the same applied voltages \cite{Martino:2016}. Figure \ref{CurrentDensity} shows the current density right after filament formation (time step (2)), which acts as an input for the temperature calculation.
The current is mainly flowing through the conductive filament locally heating up the device. Thus, the temperature increases within the set process due to Joule heating up to a value of 320.9 K. Since the current is set to the compliance current (red dotted line in figure \ref{IV}b)), the applied voltage drops (blue line in figure \ref{IV}a)). Therefore, during the current compliance the filament growth is extremely slowed. By reversing the voltage polarity, the current increases (3). Furthermore, the temperature increases to values around 305 K. As soon as the galvanic connection between the filament and the Ag electrode breaks up (4), the temperature decreases to room temperature. Due to the opposite direction of the internal electric field the filament forms back (5). Since no predefined nucleation seed is set in this approach, the nucleation at the Pt electrode occurs at random positions.  At these very positions, Ag atoms reduce preferentially and form stable cluster. Figure \ref{Field} shows the amount of the electric field at t = 0.66 s.
\begin{figure}[t!]
\centering
\includegraphics[width=3.3in]{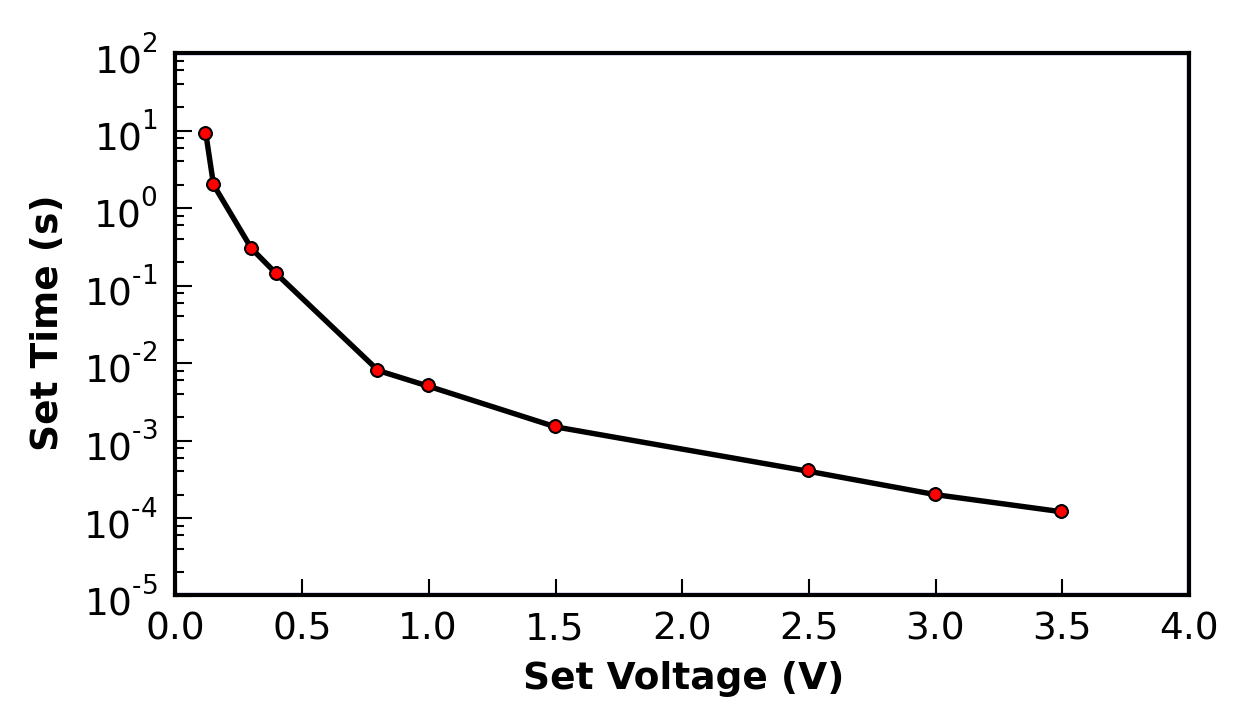}
\caption{Switching kinetic of the simulated ECM-cell.}
\label{Kinetic}
\end{figure} 

At the cluster positions the electric field is enhanced due to sharp edges and the reduced distance between the top and bottom electrode. Since Ag ions are attracted by high electric fields, filament growth is further supported at larger cluster positions. In the final analysis, filament formation occurs true to the motto: ``The winner takes it all''. The largest cluster grows most rapidly, finally forming a high conductive filament touching the Ag electrode. Consequently, since filament formation depends strongly on the electric field, the set time of ECM-cells varies with the applied voltage. In order to investigate the switching kinetic of the simulated ECM-cell, set simulations have been done for different applied voltages. In figure \ref{Kinetic} the switching kinetic of the simulated ECM-cell is shown. This switching kinetic is comparable to typical switching kinetics of ECM-cells (cf.\cite{Lübben:2017}). Of particular interest is the reset process. At the moment of maximum negative current, the calculated temperature increase is 4.6 K. Although, for higher power densities  higher temperatures within the device are possible\cite{Menzel:2013}, this simulation result shows, that for typical integrated low power devices Joule heating plays a minor role. The underlying reason for the reset process is the electric field, dismantling the conductive filament. 
\begin{figure}
\centering
\includegraphics[width=3.3in]{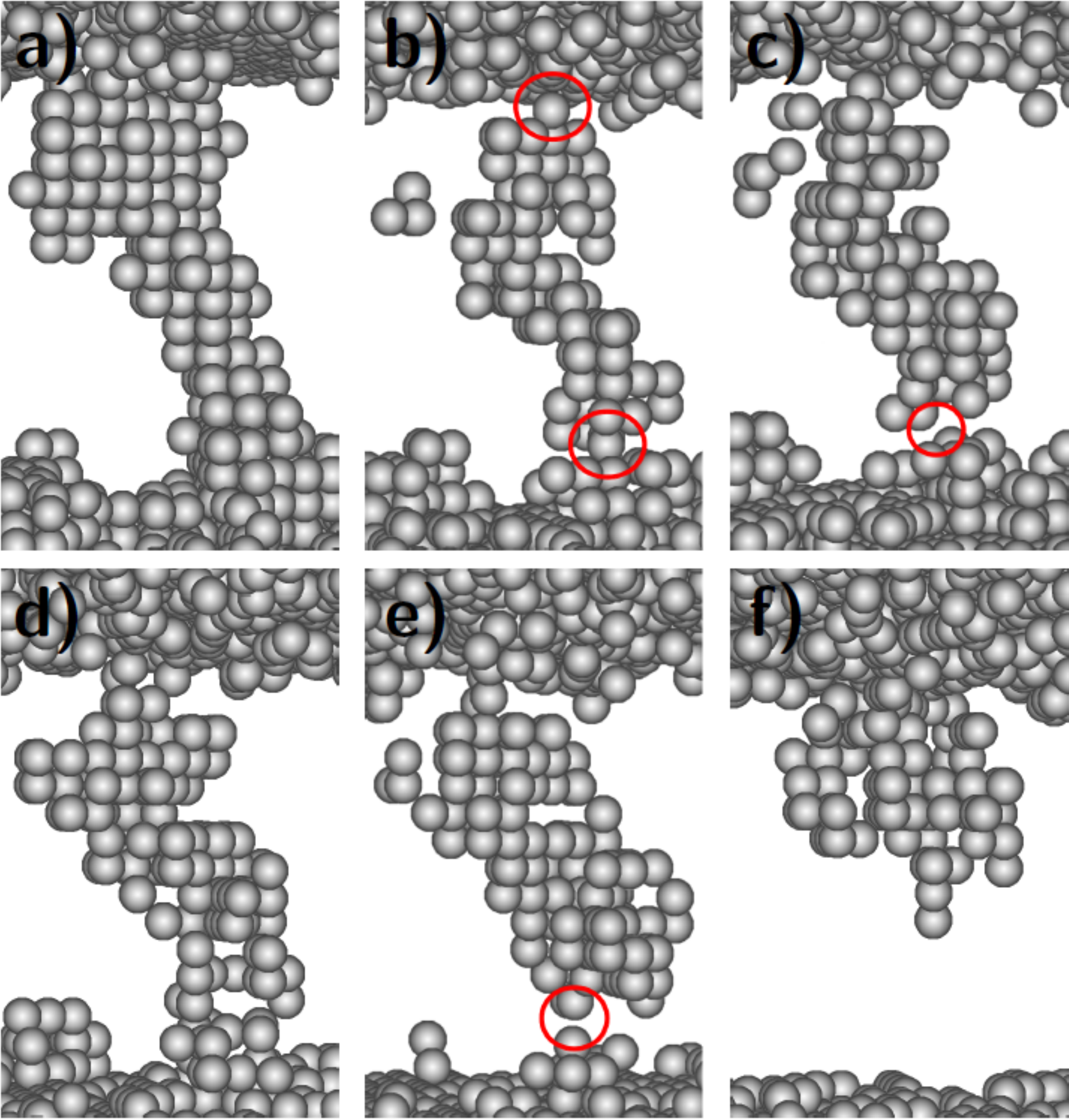}
\caption{Atomic configuration during the reset process for six selected time steps.}
\label{Reset}
\end{figure}
\begin{figure}[h!]
\includegraphics[width=3.4in]{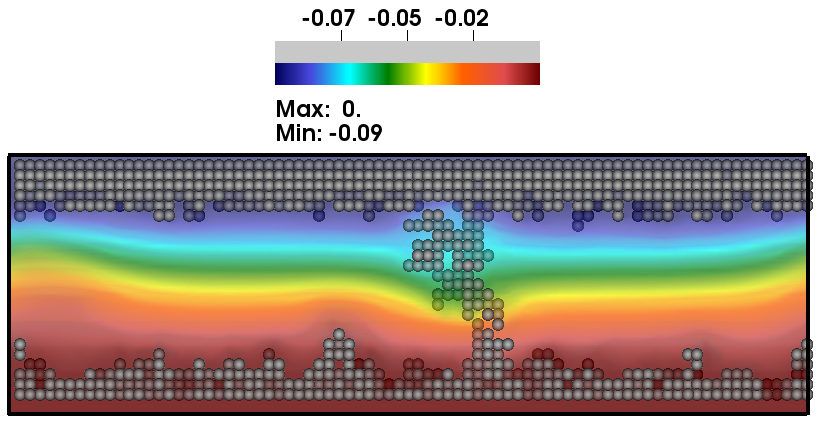}
\caption{Potential distribution (V) at t = 2.2 s right before filament rupture at a clipping through the conductive filament. \label{Potential}}
\end{figure}

In contrast to the set process, which is a relatively abrupt process, the reset process is often gradual. Furthermore, mostly the reset process is stochastic and differs from cycle to cycle\cite{Valov:2013}. In figure \ref{Reset} the atomic configuration of the conductive filament for the six selected time steps presented in figure \ref{IV}b), marked by a)-f), are shown, explaining the reset process. Figure a) shows the Ag filament before start of the reset process. Caused by the increasing voltage, Ag atoms oxidize at random positions depending on the potential drop between filament and electrolyte and remove from the filament. Figure \ref{Potential} shows the potential distribution at a clipping through the conductive filament for t = 2.2 s right before filament rupture, influencing the reset process. Due to that, an overall thinning of the filament leads to several weak connections (b)) indicated by red circles. As a result, the device resistance increases gradually, leading to the gradual and stochastic reset behavior. Preferentially, at these weak connection points the rupture of the filament occurs (c)). However, after a rupture of the filament a reconnection might occur (d)) leading to a decrease of the device resistance and therewith to an increase of the current once again. The final rupture of the filament is shown in e). During the back forming of the conductive filament, isolated islands emerge and an additional growth mode, that starts from the active top electrode, can be observed (f)). This growth mode has been recently discussed\cite{Dirkmann:2015} and has also been experimentally observed\cite{Yang:2014, Celano:2014}. 

\section{Conclusions}
We have presented a 3D simulation model for the resistive switching of ECM cells. Within the model, atomic processes have been calculated using the KMC method, whereas the electric field is calculated solving the continuity equation. The current is calculated using a generalized form of Ohm's law and the local temperature is calculated solving the heat transfer equation. Due to the 3D nature of the model, a detailed insight into atomic processes during filament growth and rupture is possible and the abrupt set as well as the gradual reset process could be explained in detail. The influence of the electric field on the bipolar resistive switching of ECM cells has been found to be much more important than the effect of an increased local temperature. We believe that with our contribution a deeper understanding of important physical and chemical processes on both, atomic length and experimental time scale during th the resistive switching of ECM cells is provided.

\section*{Acknowledgements}
The authors gratefully acknowledge financial support by the German Research Foundation (DFG) in the frame of Research Group FOR 2093.
%%%END OF MAIN TEXT%%%


%\balance


%%%REFERENCES%%%
\bibliography{rsc} %You need to replace "rsc" on this line with the name of your .bib file

\begin{thebibliography}{40}

\bibitem{Pan:2014}
F. Pan, S. Gao, C. Chen, C. Song, and F. Zeng, ``Recent progress in resistive random access memories: Materials, switching mechanisms, and performance'', \emph{Mater. Sci. Eng. R}, \textbf{83}, pp. 1 - 59, 2014.

\bibitem{Hansen:2017}
M. Hansen, F. Zahari, M. Ziegler, and H. Kohlstedt, ``Double-Barrier Memristive Devices for Unsupervised Learning and Pattern Recognition'', \emph{Front. Neurosci.}, \textbf{11}, 2017.

\bibitem{Waser:2007}
R. Waser, and M. Aono, ``Nanoionics-based resistive switching memories'', \emph{Nat. Mater.}, \textbf{6}, pp. 833 - 840, 2007.

\bibitem{Menzel:2012}
S. Menzel, U. B\"ottger, and R. Waser, ``Simulation of multilevel switching in electrochemical metallization memory cells'', \emph{J. Appl. Phys.}, \textbf{111}, pp. 014501-1 - 014501-5, 2012.

\bibitem{Yu:2011}
S. Yu, H.-S. Wong, ``Compact Modeling of Conducting-Bridge Random-Access Memory (CBRAM)'', \emph{IEEE Trans. Electron Devices}, \textbf{58}, 2011.

\bibitem{Russo:2009}
U. Russo, D. Kamalanathan, D. Ielmini, A. L. Lacaita and M. N. Kozicki, ``Study of Multilevel Programming in Programmable Metallization Cell (PMC) Memory'', \emph{IEEE Trans. Electron Devices}, \textbf{56} (5), pp. 1040 - 1047, 2009.

\bibitem{Lin:2012}
S. Lin, L. Zhao, J. Zhang, H. Wu, Y. Wang, H. Qian, and Z. Yu, ``Electrochemical Simulation of Filament Growth and Dissolution in Conductive-Bridging RAM (CBRAM) with Cylindrical Coordinates'', \emph{IEDM}, pp. 26.3.1 - 26.3.4  2012.

\bibitem{Onofrio:2015}
N. Onofrio,	D. Guzman, and A. Strachan, ``Atomic origin of ultrafast resistance switching in nanoscale electrometallization cells'', \emph{Nat. Mater.}, \textbf{14}, pp. 440 - 446, 2015.

\bibitem{Jameson:2011}
J. R. Jameson, N. Gilbert, F. Koushan, J. Saenz, J. Wang, S. Hollmer, and M. N. Kozicki, ``One-dimensional model of the programming kinetics of conductive-bridge memory cells'', \emph{Appl. Phys. Lett.}, \textbf{99}, pp. 063506-1 - 063506-3, 2011.

\bibitem{Jameson:2012}
J. R. Jameson, N. Gilbert, F. Koushan, J. Saenz, J. Wang, S. Hollmer, and M. Kozicki, ``Effects of cooperative ionic motion on programming kinetics of conductive-bridge memory cells'', \emph{Appl. Phys. Lett.}, \textbf{100}, pp. 023505-1 - 023505-4, 2012.

\bibitem{Pan:2011}
F. Pan, S. Yin, and V. Subramanian, ``A Detailed Study of the Forming Stage of an Electrochemical Resistive Switching Memory by KMC Simulation'', \emph{IEEE Electron Device Lett.}, \textbf{32}, pp. 949 - 951, 2011.

\bibitem{Qin:2015}
S. Qin, Z. Liu, G. Zhang, J. Zhang, Y. Sun, H. Wu, H. Qian, and Z. Yu, ``Atomistic study of dynamics for metallic filament growth in conductive-bridge random access memory'', \emph{Phys. Chem. Chem. Phys.}, \textbf{17}, pp. 8627 - 8632, 2015.

\bibitem{Menzel:2015}
S. Menzel, P. Kaupmann, and R. Waser, ``Understanding filamentary growth in electrochemical metallization memory cells using kinetic Monte Carlo simulations'', \emph{Nanoscale}, \textbf{7}, pp. 12673 - 12681, 2015.

\bibitem{Ielmini:2016}
D. Ielmini, and R. Waser, ``Resistive Switching: From Fundamentals of Nanoionic Redox Processes to Memristive Device Applications'', \emph{Wiley}, p. 400, 2016.

\bibitem{Dirkmann:2015}
S. Dirkmann, M. Ziegler, M. Hansen, H. Kohlstedt, J. Trieschmann, and T. Mussenbrock,
``Kinetic simulation of filament growth dynamics in memristive electrochemical metallization devices'',  \emph{J. Appl. Phys.}, \textbf{118}, pp. 214501-1 - 214501-7, 2015.

\bibitem{Enghag2004}
P. Enghag, ``Encyclopedia of the Elements: Technical Data - History - Processing - Applications'', \emph{Wiley}, p. 125, 2004.

\bibitem{Kharisov2016}
B. I. Kharisov, O. V. Kharissova, and U. Ortiz-Mendez, ``CRC Concise Encyclopedia of Nanotechnology'', \emph{CRC Press}, p. 726, 2016.

\bibitem{Abu-Eishah:2004}
S. I. Abu-Eishah, Y. Haddad, A. Solieman, and A. Bajbouj, ``A new correlation for the specific heat of metals, metal oxides and metal fluorides as a function of temperature'', \emph{Lat. Am. appl. res.}, \textbf{113} (4), pp. 257 - 264, 2004.

\bibitem{Saeedian:2013}
M. Saeedian, M. Mahjour-Shafiei, E. Shojaee, and M. R. Mohammadizadeh, ``Specific Heat Capacity of TiO$_2$ Nanoparticles'', \emph{arXiv:1307.7555}, 2013.

\bibitem{Ho:1972}
C. Y. Ho, R. W. Powell, and P. E. Liley, ``Thermal Conductivity of the Elements'', \emph{J. Phys. Chem. Ref. Data}, \textbf{1} (2), pp. 279 - 421, 1972.

\bibitem{Lu:2013}
Y. M. Lu, M. Noman, Y. N. Picard, J. A. Bain, P. A. Salvador, and M. Skowronski, ``Impact of Joule heating on the microstructure of nanoscale TiO$_2$ resistive switching devices'', \emph{J. Appl. Phys.}, \textbf{113}, pp. 163703-1 - 163703-9, 2013. 

\bibitem{Dirkmann:2016}
S. Dirkmann, M. Hansen, M. Ziegler, H. Kohlstedt, and T. Mussenbrock,
``The role of ion transport phenomena in memristive double barrier devices'',  \emph{Sci. Rep.}, \textbf{6}, 2016.

\bibitem{Yang:2011}
L. Yang, ``Resistive Switching in TiO2 Thin Films'', \emph{Forschungszentrum J\"ulich}, 2011.

\bibitem{Martino:2016}
G. Di Martino, S. Tappertzhofen, S. Hofmann, and J. Baumberg, `` Nanoscale Plasmon-Enhanced Spectroscopy in Memristive Switches'', \emph{Small}, \textbf{10}, pp. 1334 - 1341, 2016.

\bibitem{Lübben:2017}
M. L\"ubben, S. Menzel, S. G. Park, M. Yang, R. Waser, and I. Valov, ``SET kinetics of electrochemical metallization cells: influence of counter-electrodes in SiO2/Ag based systems'', \emph{Nanotechnology}, \textbf{28}, 2017.

\bibitem{Menzel:2013}
S. Menzel, I. Valov, R. Waser, N. Adler, J. van den Hurk, and S. Tappertzhofen, ``Simulation of polarity independent RESET in electrochemical metallization memory cells'', \emph{5th IEEE International Memory Workshop}, pp. 92 - 95, 2013.

\bibitem{Valov:2013}
I. Valov, E. Linn, S. Tappertzhofen, S. Schmelzer, J. van den Hurk, F. Lentz, and R. Waser, ``Nanobatteries in redox-based resistive switches require extension of memristor theory'', \emph{Nat. Commun.}, \textbf{4}, 2013.


\bibitem{Yang:2014}
Y. Yang, P. Gao, L. Li, X. Pan, S. Tappertzhofen, S. H. Choi, R. Waser, I. Valov, and W. D. Lu, ``Electrochemical dynamics of nanoscale metallic inclusions in dielectrics'', \emph{Nat. Commun.}, \textbf{5}, pp. 4232-1 - 4232-9, (2014).

\bibitem{Celano:2014}
U. Celano, L. Goux, A. Belmonte, K. Opsomer, A. Franquet, A. Schulze, C. Detavernier, O. Richard, H. Bender, M. Jurczak, and W. Vandervorst, ``Three-Dimensional Observation of the Conductive Filament in Nanoscaled Resistive Memory Devices'', \emph{Nano Lett.}, \textbf{14}, pp. 2401 - 2406, 2014.





\end{thebibliography}
\bibliographystyle{rsc} %the RSC's .bst file

\end{document}